\documentclass[twocolumn,prb,color,epsfig,superscriptaddress]{revtex4}
\usepackage[dvipdfmx]{graphicx} 

\begin{document}


\title[Local intra-unit-cell order parameters in cuprates]{Local intra-unit-cell order parameters in cuprates}

\author{A.S. Moskvin}

\address{Department of Theoretical Physics, Ural Federal University, 620083 Ekaterinburg,  Russia}

\begin{abstract}
Starting with a minimal model  with the on-site Hilbert space  reduced to only three effective valence centers CuO$_4^{7-,6-,5-}$ (nominally Cu$^{1+,2+,3+}$) we present an unified approach to the description of the variety of the local intra-unit-cell (IUC) order parameters determining a low-energy physics in cuprates. Central point of the model implies the occurrence of unconventional on-site quantum superpositions of the three valent states characterized by different hole occupation: $n_h$=0,1,2 for Cu$^{1+,2+,3+}$ centers, respectively, different conventional spin: s=1/2 for Cu$^{2+}$ center and s=0 for Cu$^{1+,3+}$ centers, and different orbital symmetry:$B_{1g}$ for the ground states  of the  Cu$^{2+}$ center and  $A_{1g}$   for the Cu$^{1+,3+}$ centers, respectively. The latter  does result in a spontaneous orbital symmetry breaking accompanying the formation of the on-site mixed valence superpositions with emergence of the IUC orbital nematic order parameter of the $B_{1g}=B_{1g}\times A_{1g}$ ($\propto d_{x2-y2}$) symmetry.
To describe the diagonal and off-diagonal, or quantum local charge order we develop an S=1 pseudospin model with a non-Heisenberg effective Hamiltonian that provides a physically clear description of "the myriad of phases"\,  from a bare parent  antiferromagnetic insulating phase to a Fermi liquid  in  overdoped  cuprates.  
Conventional spin density $\rho_s$ for mixed valence superpositions can vary inbetween 0 and 1 in accordance with the weight of the Cu$^{2+}$ center in the superposition. We show that the superconductivity and spin magnetism are nonsymbiotic phenomena with competing order parameters. 
Furthermore we argue that instead of a well-isolated Zhang-Rice (ZR) singlet $^1A_{1g}$ the ground state of the hole  Cu$^{3+}$ center in cuprates should be described by a complex $^1A_{1g}$-$^{1,3}B_{2g}$-$^{1,3}E_u$ multiplet, formed by a competition of conventional hybrid Cu 3d-O 2p $b_{1g}(\sigma)\propto d_{x^2 -y^2}$ state and {\it purely oxygen nonbonding} O 2p$\pi$ states with $a_{2g}(\pi)$ and $e_{ux,y}(\pi)$ symmetry. In contrast with inactive ZR singlet we arrive at several novel competing IUC orbital and spin-orbital order parameters, e.g., electric dipole and quadrupole moments, Ising-like net orbital magnetic moment,  orbital toroidal moment, intra-plaquette's staggered order of Ising-like oxygen orbital magnetic moments. As a most impressive validation of the non-ZR model we explain fascinating results of recent neutron scattering measurements that revealed novel type of the IUC magnetic ordering in pseudogap phase of several hole-doped cuprates.

\end{abstract}
\maketitle



\section{Introduction}

The mechanism underlying the high-temperature superconductivity of copper oxides\,\cite{Muller} has remained unelucidated and is still one of the greatest mysteries in the field of condensed matter physics. 
The cuprate high-T$_c$ superconductors start out life as antiferromagnetic insulators in contrast with BCS
superconductors being conventional metals. Unconventional behavior of these materials under charge doping, in particular, a remarkable interplay of charge, lattice, orbital, and spin degrees of freedom, strongly differs from that of ordinary metals and merely resembles that of a doped Mott insulator.
Long range antiferromagnetic (AFM) order typical for parent cuprates puzzlingly sharply disappears under doping\,\cite{Kastner}. For instance in  La$_{2-x}$Sr$_x$CuO$_4$ the AFM order disappears for $x\approx$\,0.02 when the system turns into a so-called "spin glass" phase. Superconductivity occurs beyond a minimal hole content $\delta_{min}\approx$\,0.05\,-\,0.06 where there is no long-range antiferromagnetic order.
Nevertheless, signs of magnetism persist in the hole-doped
materials with $\delta >\delta_{min}$ as a local nanoscale order, as observed in muon spin relaxation and neutron scattering measurements.
On the other hand, the heavily doped nonsuperconducting crystals are metallic and no longer exhibit the AF spin correlations; their electronic properties can be understood in terms of the normal Fermi liquid. In other words, the high-T$_c$ superconductivity appears in a doping range between the AF insulator and nominally normal Fermi liquid regions. However, in normal state, these materials exhibit non-Fermi liquid properties and enter a mysterious pseudogap (PG) regime, characterized by the observation of a multiple crossover pseudogap temperatures T$^*$'s.  In addition to the occurrence of unconventional d-wave superconductivity the phase diagram of high temperature superconducting (HTSC) copper oxides does reveal a flurry of various anomalous electronic properties.  The analysis of the PG phase in underdoped cuprates remains a hot topic in research on correlated electron systems.
A large variety of intensive experimental studies of underdoped cuprates provide direct or indirect indications that a symmetry breaking state develops below the PG temperature, supporting the competing order scenario in the sense that the data show that the PG is likely a phase (or even a set of phases) with broken symmetries.

The evidence for broken symmetries comes from different sources. The presence of stripe-like charge and spin density ordering has been known for a long time to exist in 214 compounds\,\cite{stripe}. Charge modulated structure in Bi$_2$Sr$_2$CaCu$_2$O$_8$ (BSCCO) was observed by STM technique\,\cite{Hoffman}. Most notably, recent nuclear magnetic resonance (NMR)\,\cite{CDW-NMR}, scanning tunneling microscopy (STM)\,\cite{STM}, x-ray scattering (XRS)\,\cite{CDW-XRS},  and other measurements on a number of different families of the hole-doped cuprates have detected the onset of an incommensurate charge density wave (CDW) state in the CuO$_2$ plane at a temperature T$_{CDW}$, T$_c$\,$<$\,T$_{CDW}$\,$<$\,T$^*$, which are stabilized to static, long range order by a magnetic field, and in general compete with superconductivity.
Novel CDW states differ in some important respects from the more familiar stripes observed in the 214 family, in particular, the doping dependence of the in-plane wave-vectors in the two systems is quite different.  
There is growing experimental evidence that charge density wave order is a generic feature   and  an important competitor to superconductivity in underdoped cuprate high temperature superconductors.
Understanding the properties of these charge ordered phases, competing or coexisting with superconductivity, may significantly help to clarify the physical origin of the PG phase.

A significant property of the charge order in cuprates is that it can exhibit both inter- and intra-unit cell symmetry breaking. 
In recent years, there has been increasing experimental evidence pointing towards the existence of an electronic nematic phase in some high-Tc superconductors. Intra-unit-cell (IUC) nematicity, or the breaking of rotational symmetry by the electronic structure within each CuO$_2$ unit cell is often observed in STM measurements in cuprates\,\cite{nematic} and interpreted as a charge density inequivalence for two pairs of oxygen ions oriented along  $a$- and $b$-axes, respectively.

Two sets of experiments indicate that time reversal symmetry may be also broken in the pseudogap phase of  cuprates. One is the observation of the polar Kerr effect in YBa$_2$Cu$_3$O$_{6+x}$ (YBCO) and La$_{1.875}$Ba$_{0.125}$CuO$_4$ (LBCO) below some critical temperature T$_K$(x), which increases as $x$ decreases\,\cite{Kerr}. Another is the observation of intra-unit-cell (IUC) magnetic order in polarized neutron scattering measurements\,\cite{Bourges}. The puzzling IUC order breaks time reversal symmetry, but preserves lattice translation invariance. However, at variance with ferromagnets, it does not give rise to a uniform magnetization.
It is worth noting that together with persistent evidence of other forms of electronic order in hole-doped cuprates different local probes have provided numerous examples where superconducting, electronic, and magnetic properties vary on a nanoscopic length scale.
A large variety of theoretical models has been designed to account for these exotic electronic properties and to shed light on their interplay with nonconventional superconductivity. However, the most important questions, particularly on the microscopic structure of the order parameters, remain unanswered to date. 

In this paper, we consider a simplified minimal microscopic model of the CuO$_4$ centers in cuprates which allows to derive all the IUC order parameters in frames of an unified approach and show that all the main features of superconductivity, charge order, including charge-nematicity, spin and orbital order can be explained on equal footing. 
The rest of the paper is organized as follows. In Section II we introduce a physical working model of the CuO$_4$ centers in cuprates, first an S=1 pseudospin formalism to describe local IUC quantum charge order parameters. Section III contains a brief overview of the non-Zhang-Rice model for the nominally Cu$^{3+}$ centers in cuprates, characterization of novel IUC spin-orbital order parameters, and their manifestations in recent experimental data. A short conclusion is made in Sec.IV.

\section{Simple toy model of the alternating valence system and the disproportionation driven superconductivity}

\subsection{Pseudospin formalism}

Valent electronic states in  strongly correlated 3\emph{d} oxides manifest
both significant correlations and \emph{p}-\emph{d} covalency with a distinct trend to localisation of many-electron configurations formed by antibonding Me 3d-O 2p hybridized molecular orbitals. The localisation effects are particularly clear featured in the crystal field \emph{d}-\emph{d} transitions whose spectra just weakly vary from diluted to concentrated 3\emph{d} oxides. An optimal way to describe valent electronic states in  strongly correlated 3\emph{d} oxides is provided by quantum-chemical techniques such as the ligand field theory which implies a crystal composed of a system of small 3d-cation-anion clusters. Naturally, such an approach has a number of 
shortcomings, nevertheless, this provides a clear physical
picture of the complex electronic structure and the energy
spectrum, as well as the possibility of a quantitative modelling.
In a certain sense the cluster calculations might provide a better description of the overall
electronic structure of  insulating  3\emph{d} oxides  than different so-called {\it ab-initio} 
band structure calculations, mainly  due to a better account for correlation effects and electron-lattice coupling. 


Given the complexity of the inter-related charge deformation dynamics in cuprates with the disproportionation instability\,\cite{Moskvin-11}, we introduce a minimal model to describe its low-energy physics with the on-site Hilbert space  reduced to ground states of only three  effective valence centers CuO$_4^{7-,6-,5-}$ (nominally Cu$^{1+,2+,3+}$) where the electronic and lattice degrees of freedom get strongly locked together. Despite some simplifications such a charge triplet's lattice model with the three valence states of the CuO$_4$ center described on an equal footing is believed to capture the salient features both of the hole- and electron-doped cuprates\,\cite{SCES}.
  
\begin{figure}[t]
\includegraphics[width=8.5cm,angle=0]{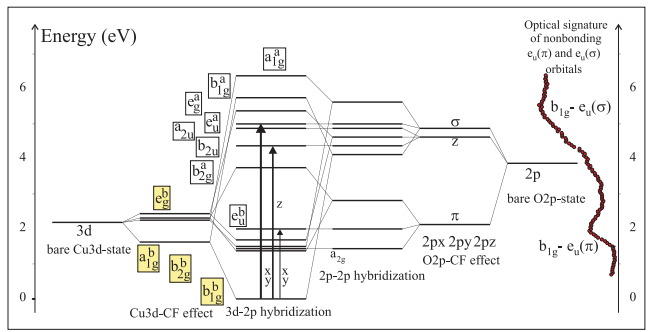}
\caption{(Color online) Model single-hole energy spectrum for a CuO$_4$ plaquette
with parameters relevant for Sr$_2$CuO$_2$Cl$_2$ and a number of
other insulating cuprates with corner-shared CuO$_4$ plaquettes. Also shown is a fragment of the EELS spectrum for Sr$_2$CuO$_3$\,\cite{Moskvin-PRL} with a distinct signature of the two main planar (${\bf E}\perp C_4$) dipole-allowed \emph{p-d} CT transitions $b_{1g}\rightarrow e_u(\pi)$ and $b_{1g}\rightarrow e_u(\sigma)$, respectively.
} \label{fig1}
\end{figure}

Validity of such an approach implies well isolated ground states of the three centers. This surely holds for the $^1A_{1g}$ singlet ground state of the Cu$^{1+}$ centers with nominally filled 3d shell whose excitation energy does usually exceed 2\,eV (see, e.g., Ref.\,\onlinecite{Pisarev-2006} and references therein). 
Fig.\,\ref{fig1}  presents a  single-hole energy spectrum for a CuO$_4$ plaquette, or Cu$^{2+}$ center, embedded into an insulating cuprate such as Sr$_2$CuO$_2$Cl$_2$ calculated with a reasonable set of parameters\,\cite{Moskvin-PRB-02}. For illustration we show  a  step-by-step formation of the cluster energy levels from the bare Cu 3d and O 2p levels with the successive inclusion of crystalline field (CF) effects, O 2p - O 2p, and Cu 3d - O 2p covalency. It is worth noting that the $b_{1g}\propto d_{x^2-y^2}$ character of the ground hole state in CuO$_4^{6-}$ cluster seems to be one of a few indisputable points in cuprate physics. A set of low-lying excited states with the energy $\geq$\,1.5\,eV includes bonding molecular orbitals with $a_{1g}\propto d_{z^2}$, $b_{2g}\propto d_{xy}$, and $e_{g}\propto d_{yz},d_{xz}$ symmetry, as well as purely oxygen nonbonding orbitals with $a_{2g}(\pi)$ and $e_u(\pi)$ symmetry. The results agree with experimental data  obtained by optical, EELS, and RIXS technique  for different cuprates (see, e.g., Refs.\,\onlinecite{Moskvin-PRB-02,Moskvin-PRL} and Fig.\,\ref{fig1}).

In 1988 Zhang and Rice\,\cite{ZR} have proposed that the doped hole in parent cuprate forms a Cu$^{3+}$ center with a well isolated local spin and orbital ${}^1A_{1g}$ singlet ground state which involves a phase coherent combination of the 2p$\sigma$ orbitals of the four nearest neighbor oxygens with the same $b_{1g}$ symmetry as for a bare Cu 3$d_{x^2-y^2}$ hole.  Here, it should be noted that
when speaking of a Zhang-Rice singlet as being "well isolated", one implies that the ${}^{1}A_{1g}$ ground state for the CuO$_4$ plaquette with the two holes of the $b_{1g}(d_{x^2-y^2})$ symmetry is well separated by more than 1\,eV from any other excited two-hole states. From the very beginning of 90ths the Zhang-Rice (ZR)  model despite a lack of straightforward experimental evidence becomes a mainstream theoretical approach to low-energy cuprate physics. 




One strategy to cast into a tractable model the physics of the charge triplets is to make use of a S=1 pseudospin formalism and to create model pseudospin Hamiltonian which can reasonably well reproduce both the ground state and important low-energy excitations of the full problem. Standard pseudospin formalism represents a variant of the equivalent operators technique widely known in different physical problems from classical and quantum lattice gases, binary alloys, (anti)ferroelectrics,.. to neural networks\,\cite{Batista+Ortiz}. The formalism starts with a finite basis set for a lattice site (triplet of $M^0,M^{\pm}$ centers in our model, see below). Such an approach differs from well-known pseudospin-particle transformations akin Jordan-Wigner\,\cite{JW} or Holstein-Primakoff\,\cite{HP} transformation which establish a strict linkage between pseudospin operators and the creation/annihilation operators of the Fermi or Bose type. The pseudospin formalism for electron systems generally proceeds with a truncated basis and does not imply a strict relation to fermion operators that obey the fermionic anti-commutation rules. 
It is worth noting that all formulations of superconductivity are reduced to a pairing instability of well-defined quasiparticles.
The identifying the weakly interacting entities that make a particle interpretation of the current possible became one of the key problem that arises from the strong correlations in the normal state of the copper-oxide superconductors.  However, there is good reason to believe that the construction of such entities may not be possible\,\cite{Phillips-2013}.

Generally speaking, at variance with conventional Hubbard models  the charge pseudospin system cannot be ascribed neither to conventional Fermi nor to Bose systems similarly to a familiar slave-boson system\,\cite{KR}.  
In the Kotliar-Ruckenstein slave-boson formalism\,\cite{KR}, the local Hilbert space of the
Hubbard model is expanded by introducing a fermion ${\hat f}_{\sigma}$, which stands
for the $\sigma$-spin QP and one slave boson
for each Fock state as ${\hat e}$ for the empty state (holon) $|0\rangle$, ${\hat p}_{\sigma}$
for the singly occupied state $|\sigma\rangle$($\sigma$\,=\,$\uparrow$ or $\downarrow$), and ${\hat d}$ for the doubly occupied state (doublon) $|\uparrow\downarrow\rangle$. The three Fock states can be addressed to be the local Hilbert space of the semi-hard core bosons which 
can be mapped into a system of S = 1 centers via a generalization of the Matsubara-Matsuda transformation\,\cite{MM} (see Ref.\,\cite{EBHM} and references therein) that also maps the boson density into  the local magnetization: $n_j = S_{zj} + 1$. In contrast to the hard-core bosons associated with S = 1/2 magnets, this makes possible to study "Hubbard-like"\, bosonic gases with on-site density-density interactions because $n_j \leq 2$.
It is worth noting that the system of  charge  triplets and the S=1 pseudo-spin formalism were used earlier to  describe  the neutral-to-ionic electronic-structural transformation   in organic charge-transfer crystals (see, e.g., paper by T. Luty in Ref.\,\onlinecite{Toyozawa}).


Hereafter we address a simplified toy  model of a mixed-valence system with three possible stable valence states of a cation-anion
 cluster CuO$_4$ ($M$: $M^0, M^{\pm}$), forming the charge (isospin) triplet and neglect all other degrees of freedom focusing on the quantum charges\,\cite{Moskvin-LTP,Moskvin-11,SCES}. 
  Three different valence charge states of the $M$-center:
  $M^0,M^{\pm}$ we associate   with three components of the $S=1$ pseudo-spin (isospin)
triplet with pseudospin projections  $M_S =0,+1,-1$, respectively. Having in mind quasi-2D cuprates we associate 
$M^0,M^{\pm}$ centers with three charge states of the CuO$_4$ plaquette: a bare center $M^0$=CuO$_4^{6-}$, a hole center $M^{+}$=CuO$_4^{5-}$, and an electron center $M^{-}$=CuO$_4^{7-}$, respectively. However, we should note that at variance with spinless ground states of Cu$^{1+}$  and Cu$^{3+}$ centers the bare Cu$^{2+}$ center has conventional spin s=1/2, in other words we arrive at the S\,=\,1 pseudospin model with doubly degenerate M=0 value\,\cite{Borgs,Micnas-JPCM-2012}.
In the partition function of the classical spin model, this leads to a factor of 2 for every Cu$^{2+}$ site.
 

The $S=1$ spin algebra includes three independent irreducible tensors
${\hat V}^{k}_{q}$ of rank $k=0,1,2$ with one, three, and five components,
respectively, obeying the Wigner-Eckart theorem \cite{Varshalovich}
\begin{equation}
\langle SM| {\hat V}^{k}_{q}| SM^{'}\rangle=(-1)^{S-M} \left(
\begin{array}{ccc}S&k&S
\\
-M&q&M^{'}\end{array}\right)\left \langle S\right\| {\hat
V}^{k}\left\|S\right\rangle. \label{matelem}
\end{equation}
Here we make use of standard symbols for the Wigner coefficients and reduced
matrix elements. In a more conventional Cartesian scheme a complete set of the
non-trivial pseudo-spin operators would include both ${\bf   S}$ and a number
of symmetrized bilinear forms $\{S_{i}S_{j}\}=(S_{i}S_{j}+S_{j}S_{i})$, or
spin-quadrupole operators, which are linearly coupled to $V^{1}_{q}$ and $V^
{2}_{q}$, respectively
$$
V^{1}_{q}=S_{q}; S_{0}=S_{z}, S_{\pm}=\mp \frac{1}{\sqrt{2}}(S_{x}\pm iS_{y} ):
$$
\begin{equation}
V^{2}_{0} \propto (3S_{z}^{2}-{\bf  S}^2), V^{2}_{\pm 1}\propto (S_z S_{\pm}+
S_{\pm}S_z), V^{2}_{\pm 2}\propto S_{\pm}^2 .
\end{equation}

Instead of the three $|1M\rangle $ states one may use the Cartesian basis set ${\bf \Psi}$, or $|x,y,z\rangle$:
\begin{equation}
	|10\rangle = |z\rangle\,,|1\pm 1\rangle = \mp\frac{1}{\sqrt{2}}(|x\rangle \pm i|y\rangle ) 
\end{equation}
The pseudospin matrix  has   a very simple form within the $|x,y,z\rangle$ basis set:
\begin{eqnarray}
\langle i |\hat S_{k} | j \rangle =i \epsilon _{ikj}.
\end{eqnarray}

We start by introducing the following set of S=1 coherent states characterized by vectors $\bf a$ and $\bf b$ satisfying the normalization constraint\,\cite{Knig,Nadya} 
\begin{eqnarray}
|{\bf c}\rangle =|{\bf a}, {\bf b}\rangle = {\bf c}\cdot{\bf \Psi}=({\bf a} +i{\bf b})\cdot{\bf \Psi}
\label{ab}
\end{eqnarray}
where ${\bf a}$ and ${\bf b}$ are real vectors that are
arbitrarily oriented with respect to some fixed coordinate
system in the pseudospin space with orthonormal basis ${\bf e}_{1,2,3}$. 

The two vectors are coupled, so the minimal number of
dynamic variables describing the $S=1$ (pseudo)spin system appears to be equal to four.
Hereafter we would like to emphasize the $director$ nature of the ${\bf c}$
vector field: $|{\bf c}\rangle$ and $|-{\bf c}\rangle$ describe the physically
identical states.

It should be noted that in a real space the $|{\bf c}\rangle$ state corresponds to a quantum on-site superposition 
\begin{eqnarray}
|{\bf c}\rangle = c_{-1}|Cu^{1+}\rangle +c_0|Cu^{2+}\rangle +c_{+1}|Cu^{3+}\rangle \, .
\label{function}
\end{eqnarray}
Existence of such unconventional on-site superpositions is a princial point of our model.

Below instead of $\bf a$ and $\bf b$ we will make use of a pair of unit vectors $\bf m$ and $\bf n$, defined as follows\,\cite{Knig}: 
$$\bf a\,=\,\cos\varphi\,\,{\bf m} ;\,\, \bf b\,=\,\sin\varphi\,\,{\bf n}\,.
$$

 For the averages of the principal pseudospin operators we obtain
$$
\langle {\bf S} \rangle = \sin2\varphi[{\bf m} \times {\bf n}],
$$
\begin{eqnarray}
\langle\{S_{i},S_{j}\}\rangle =2(\delta_{ij}-\cos^2\varphi \,m_{i}m_{j}-\sin^2\varphi \,n_{i}n_{j})\, ,
\end{eqnarray}
or
$$
\langle S_{i}^2\rangle =1-\frac{1}{2}(m_{i}^2+n_{i}^2)-\frac{1}{2}(m_{i}^2-n_{i}^2)\cos2\varphi \, , 
$$
$$
\langle\{S_{i},S_{j}\}\rangle =-(m_{i}m_{j}+n_{i}n_{j})-
$$
\begin{equation}
	(m_{i}m_{j}-n_{i}n_{j})\cos2\varphi\,, (i\not= j) \,.
\end{equation}

One should note a principal difference between the $S=\frac{1}{2}$ and  $S=1$ quantum systems. The only on-site order parameter in the former case is an average spin moment $\langle S_{x,y,z}\rangle$, whereas in the latter one has five additional "spin-quadrupole", or spin-nematic order parameters described by a traceless symmetric tensor
\begin{equation}
	Q_{ij}=\langle (\frac{1}{2}\{S_{i},S_{j}\}-\frac{2}{3}\delta_{ij})\rangle .
\end{equation}
 Interestingly, that in a sense, the  $S=\frac{1}{2}$ quantum spin system is closer to a classic one  ($S\rightarrow \infty$) with all the order parameters defined by a simple on-site vectorial order parameter $\langle {\bf S}\rangle$  than the  $S=1$ quantum spin system with its eight independent on-site  order parameters.


\begin{figure*}[t]
\begin{center}
\includegraphics[width=14cm,angle=0]{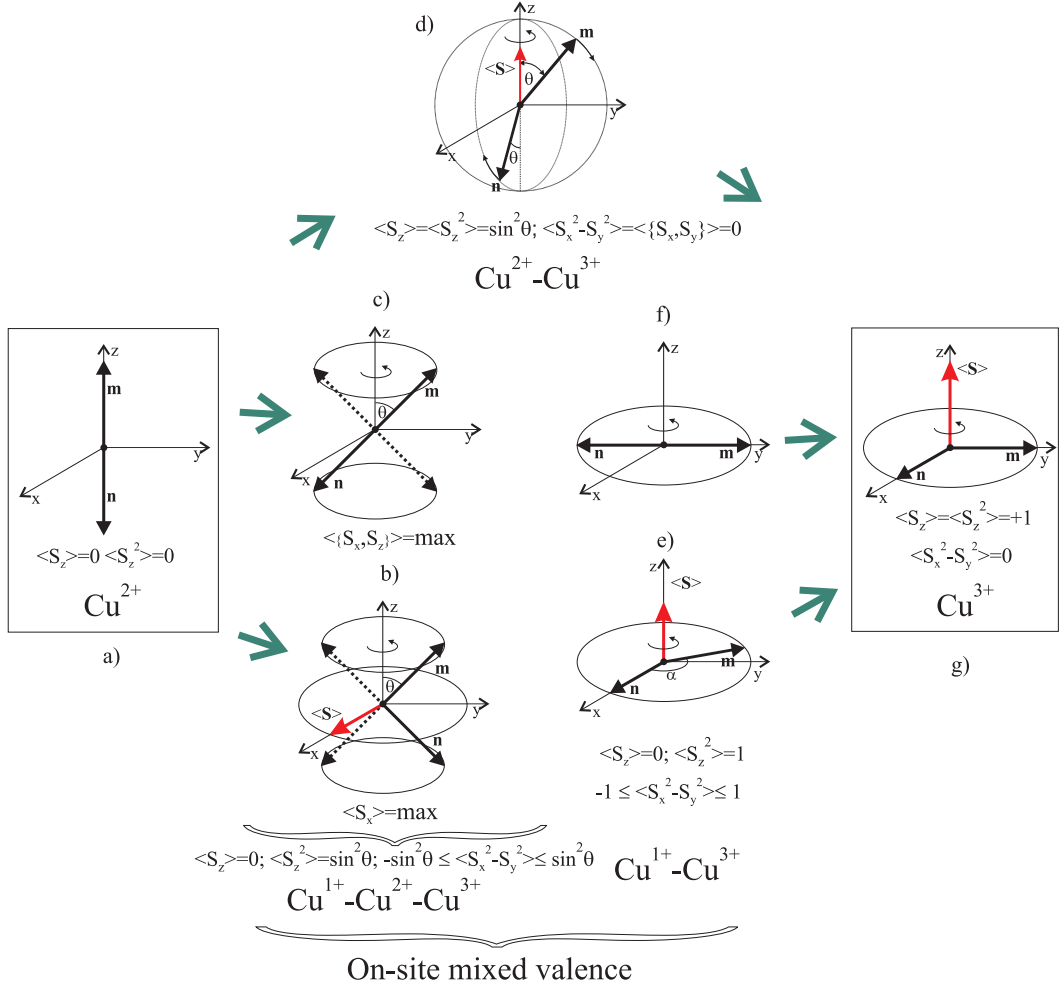}
\caption{(Color online) Cartoon showing orientations of the ${\bf m}$ and  ${\bf n}$ vectors which provide extremal values of different on-site pseudospin order parameters given $\varphi = \pi /4$ (see text for more detail).} \label{fig2}
\end{center}
\end{figure*}

To describe different types of  pseudo-spin ordering in a mixed-valence system
we make use of the eight local (on-site) order parameters: two classical ($diagonal$) order parameters: $\langle S_z
\rangle$ being a "valence", or charge density with an electro-neutrality
constraint, and $\langle S_{z}^{2} \rangle$ being the density of polar centers $M^{\pm}$, or "ionicity", and six {\it off-diagonal}
order parameters $\langle V^{k}_{q}\rangle$ ($q\not=0$). 
The {\it
off-diagonal} order parameters describe different types of the valence mixing.
Indeed, operators $V^{k}_{q}$ ($q\not=0$) change the $z$-projection of the
pseudo-spin and transform the $|SM_S\rangle$ state into  $|SM_{S}+q\rangle$
one. In other words, these can change both the $valence$ and $ionicity$.
It should be emphasized that for the $S=1$ pseudospin algebra there are two operators: $V^{1}_{\pm 1}$ and $V^{2}_{\pm 1}$, or $S_{\pm}$ and $T_{\pm}=\{S_z, S_{\pm}\}$ that change the pseudo-spin projection by $\pm 1$, with slightly different properties
\begin{equation}
\langle 0 |\hat S_{\pm} | \mp 1 \rangle = \langle \pm 1 |\hat S_{\pm} | 0
\rangle =\mp 1, \label{S1}
\end{equation}
but
\begin{equation}
\langle 0 |\hat T_{\pm}| \mp 1 \rangle = -\langle \pm 1 |(\hat T_{\pm}| 0 \rangle =+1. \label{S2}
\end{equation}
It is worth noting the similar behavior of the both operators under the hermitian conjugation:
${\hat S}_{\pm}^{\dagger}=-{\hat S}_{\mp}$;  ${\hat T}_{\pm}^{\dagger}=-{\hat T}_{\mp}$.

The $V^{2}_{\pm 2}$, or ${\hat S}_{\pm}^{2}$ 
operator changes the pseudo-spin projection by $\pm 2$ with the local order parameter 
$$
\langle S_{\pm}^{2} \rangle\,=\,\frac{1}{2}(\langle S_x^2-S_y^2\rangle \pm i\langle\{S_x,S_y\}\rangle )=
$$
\begin{equation}
c_+^*c_-=c_x^2-c_y^2\pm 2ic_xc_y\, .
\end{equation}
Obviously, this on-site off-diagonal order parameter is nonzero only when both $c_+$ and $c_-$ are nonzero, or for the on-site $M^-$(Cu$^{1+}$)-$M^+$(Cu$^{3+}$) superpositions. It is worth noting that the ${\hat S}_{+}^{2}$ (${\hat S}_{-}^{2}$) operator
creates an on-site hole (electron) pair, or composite boson, with a kinematic constraint $({\hat S}_{\pm}^{2})^2$\,=\,0, that underlines its "hard-core"\, nature.


Both ${\hat S}_{+}$(${\hat S}_{-}$) and ${\hat T}_{+}$(${\hat T}_{-}$) can be associated with the single particle creation (annihilation) operators, however, these are not standard fermionic ones, as well as ${\hat S}_{+}^2$(${\hat S}_{-}^2$) operators are not standard bosonic ones. Nevertheless, namely $\langle S_{\pm}^{2} \rangle$ can be addressed as a local superconducting order parameter

Fig.\,\ref{fig2} shows orientations of the ${\bf m}$ and  ${\bf n}$ vectors which provide extremal values of different on-site pseudospin order parameters given $\varphi = \pi /4$. The monovalent Cu$^{2+}$, or $M^0$ center, is described by a pair of $\bf m$ and $\bf n$ vectors directed along Z-axis with $|m_z|=|n_z|$\,=\,1. We arrive at the  Cu$^{2+}$-Cu$^{3+}$ ($M^0$-$M^+$) or Cu$^{2+}$-Cu$^{1+}$ ($M^0$-$M^-$) mixtures if turn $c_{-1}$ or $c_{+1}$, respectively, into zero. The mixtures  are described by a pair of $\bf m$ and $\bf n$ vectors whose projections on the XY-plane, ${\bf m}_{\perp}$ and ${\bf n}_{\perp}$, are of the same length and orthogonal to each other:${\bf m}_{\perp}\cdot{\bf n}_{\perp}$\,=\,0, $m_{\perp}$\,=\,$n_{\perp}$ with $[{\bf m}_{\perp}\times{\bf n}_{\perp}]$\,=\,$\langle S_z\rangle$\,=\,$\pm \sin^2\theta$ for $M^0$-$M^{\pm}$ mixtures, respectively (see Fig.\ref{fig2}). 

It is worth noting that for "conical"\, configurations in Figs.\,2b-2d:
$$
\langle S_z\rangle=0;\, \langle S_z^2\rangle=\sin^2\theta;\,\langle S_{\pm}^2\rangle=- \frac{1}{2}\sin^2\theta \,e^{\pm 2i\varphi}
$$
\begin{equation}
\langle S_{\pm}\rangle=-\frac{i}{\sqrt{2}}\sin2\theta \,e^{\pm i\varphi};\, \langle T_{\pm}\rangle=0 \, ,
\end{equation}
(Fig.\,2b)
$$
\langle S_z\rangle=0;\, \langle S_z^2\rangle=\sin^2\theta;\,\langle S_{\pm}^2\rangle=- \frac{1}{2}\sin^2\theta \,e^{\pm 2i\varphi}
$$
\begin{equation}
\langle S_{\pm}\rangle=0;\,\langle T_{\pm}\rangle=\mp \frac{1}{\sqrt{2}}\sin2\theta \,e^{\pm i\varphi};\, 
\end{equation}
(Fig.\,2c)
$$
\langle S_z\rangle=-\langle S_z^2\rangle=-\sin^2\theta;\,\langle S_{\pm}^2\rangle=0
$$
\begin{equation}
\langle S_{\pm}\rangle=\langle T_{\pm}\rangle=\pm \frac{1}{2}e^{\mp i\frac{\pi}{4}}\sin2\theta \,e^{\pm i\varphi}\, ,
\end{equation}
(Fig.\,2d). Figures 2e,f do show the orientation of ${\bf m}$ and ${\bf n}$ vectors for the local binary mixture Cu$^{1+}$-Cu$^{3+}$, and Fig.2g does for monovalent Cu$^{3+}$ center.
It is worth noting that for binary mixtures Cu$^{1+}$-Cu$^{2+}$ and Cu$^{3+}$-Cu$^{2+}$ we arrive at the same algebra of the ${\hat S}_{\pm}$ and ${\hat T}_{\pm}$ operators with $\langle S_{\pm}\rangle=\langle T_{\pm}\rangle$, while for ternary mixtures Cu$^{1+}$-Cu$^{2+}$-Cu$^{3+}$ these operators describe different excitations.
Interestingly that in all the cases the local Cu$^{2+}$ fraction can be written as follows:
\begin{equation}
	\rho (Cu^{2+})=1-\langle S_z^2\rangle = \cos^2\theta \, .
\end{equation}

\subsection{Orbital degree of freedom and intra-plaquette charge nematicity}

\begin{figure}[t]
\includegraphics[width=6.5cm,angle=0]{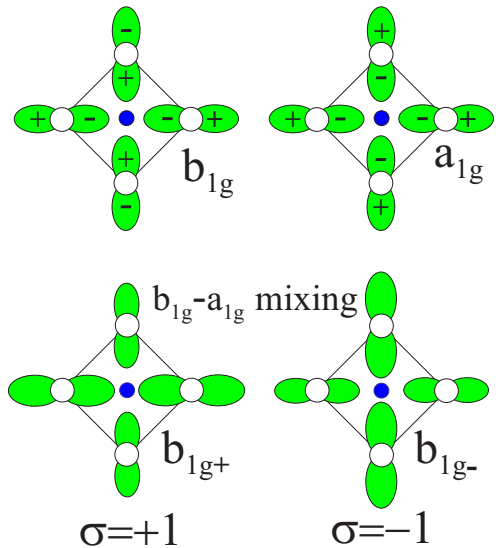}
\caption{(Color online) Top panel: oxygen charge density distribution for $b_{1g}$ and $a_{1g}$ molecular orbitals. Bottom panel: oxygen charge density distribution for two nematic states.} \label{fig3}
\end{figure}

Different orbital symmetry, $B_{1g}$ and  $A_{1g}$   of the ground states on the one side for  Cu$^{2+}$ and for Cu$^{1+,3+}$ on the other side, respectively, unequivocally should result in a spontaneous orbital symmetry breaking accompanying the formation of the on-site mixed valence superpositions (\ref{function}) with emergence of the on-site orbital order parameter of the $B_{1g}=B_{1g}\times A_{1g}$ ($\propto d_{x^2-y^2}$) symmetry. 
In frames of the CuO$_4$ cluster model the rhombic $B_{1g}$-type  symmetry breaking may be realized both by the $b_{1g}$-$a_{1g}$ ($d_{x^2-y^2}$-$d_{z^2}$) mixing for central Cu ion or  through the  oxygen subsystem either by emergence of different charge densities on the oxygens placed symmetrically relative to the central Cu ion (see Fig.\,\ref{fig3}) and/or by the $B_{1g}$-type distortion of the CuO$_4$ plaquette resulting in different Cu-O separations for these oxygens.  The latter effect seems to be natural for Cu$^{1+}$ admixtures. Indeed, at variance with Cu$^{2+}$ and Cu$^{3+}$ ions the Cu$^{1+}$ ion due to a large intra-atomic $s$\,-\,$d$\,-\,hybridization does  prefer a dumbbell O-Cu-O linear configuration thus making large rhombic distortions of the CuO$_4$ cluster.  Our effective model does not consider the cluster distortions and cannot provide detailed microscopic description of the symmetry breaking effects. Nevertheless, in frames of our effective CuO$_4$ cluster model the $B_{1g}$-type orbital symmetry breaking can be described e.g. by a simple transformation of the single particle molecular orbital
$$
|b_{1g}\rangle\rightarrow  c_b|b_{1g}\rangle +\sigma \rho^{nem} c_a|a_{1g}\rangle,
$$
where $|a_{1g}\rangle$ is the fully symmetric superposition of the four oxygen O2p$\sigma$ orbitals (see Fig.\,\ref{fig3}), $\sigma$=$\pm 1$ is a dichotomic nematic variable, and $\rho^{nem}$\,=\,$\langle S_z^2\rangle$ is a local nematic density defined as a weight of the polar centers Cu$^{1+}$ and Cu$^{3+}$ in the superposition (\ref{function}). In other words, this implies the same symmetry breaking effect for these two different polar centers. The transformation gives rise to a breaking of the four-fold rotational symmetry with nonequivalence of the charge density on the oxygen ions representing intra-unit-cell nematicity: the breaking of rotational symmetry $C_4\rightarrow C_2$ within the CuO$_4$ plaquette.
STM measurements of a static nematic order in cuprates\,\cite{nematic} support a charge imbalance between the density of holes at the oxygen sites  oriented along  $a$- and $b$-axes, however, there are clear signatures of the $B_{1g}$-type distortion (half-breathing mode) instabilities even in hole-doped superconducting cuprates which can be addressed to be a true "smoking gun"\,for electronic Cu$^{3+}$ centers. 
Indeed, two dynamically coexisting sets of CuO$_4$ clusters with different in-plane Cu-O interatomic distances have been really found by polarized Cu K-edge EXAFS in La$_{1.85}$Sr$_{0.15}$CuO$_4$\,\cite{Bianconi}. Giant phonon softening and line broadening of electronic origin of the longitudinal Cu-O bond stretching phonons near half-way to the zone boundary was observed in hole-doped cuprates (see, e.g., Ref.\onlinecite{phonon} and references therein).Their amplitude follows the superconducting dome that support our message about a specific role of electron-hole Cu$^{1+}$-Cu$^{3+}$ pairs in high-T$_c$ superconductivity\,\cite{Moskvin-11,SCES}.


\subsection{Conventional spin degree of freedom}

In the above discussion we addressed only charge and orbital degrees of freedom and neglected conventional spin s=1/2 for the Cu$^{2+}$ centers.
However, numerous experimental data, firstly that of magnetic susceptibility\,\cite{Johnston}, $\mu$-meson spectroscopy\,\cite{mu}, and magnetic neutron scattering\,\cite{neutron}, unambiguously point to a manifestation of a spin magnetism all over the phase diagram of doped cuprates.

First of all we should note that  $\hat{\rho}^s=(1-{\hat S}_{z}^2)$ to be a projection operator which picks out the s=1/2 Cu$^{2+}$ center in the on-site mixed valence superpositions (\ref{function}) can be addressed to be an on-site spin density operator. 

Conventional spin degree of freedom can be build in our effective Hamiltonian, if we transform a conventional Heisenberg spin exchange Cu$^{2+}$-Cu$^{2+}$ coupling as follows
\begin{equation}
	{\hat H}_{ex}=\sum_{i>j}{\hat I}_{ij}(\hat {\bf s}_i\cdot \hat {\bf s}_j)\, ,
	\label{Hspin}
\end{equation}
where instead of  the conventional Cu$^{2+}$-Cu$^{2+}$ exchange integral $I_{ij}$ we arrive at an effective pseudospin operator
\begin{equation}
	{\hat I}_{ij}=\hat{\rho}^s_iI_{ij}\hat{\rho}^s_j=(1-{\hat S}_{iz}^2)I_{ij}(1-{\hat S}_{jz}^2) \, .
	\label{I}
\end{equation}
In general, the isotropic bilinear exchange spin Hamiltonian in (\ref{Hspin}) should be supplied by a relatively small symmetric exchange easy-plane  anisotropy and antisymmetric Dzialoshinskii-Moriya coupling\,\cite{Kastner,DM} as follows:
\begin{equation}
	I_{ij}(\hat {\bf s}_i\cdot \hat {\bf s}_j)+K_{ij}{\hat s}_{iz}{\hat s}_{jz}+{\bf d}_{ij}[\hat {\bf s}_i\times \hat {\bf s}_j]\, .
\end{equation}

Obviously, the spin exchange provides an energy gain to the parent antiferromagnetic insulating (AFMI) phase with $\langle {\hat S}_{iz}^2\rangle$\,=\,0, while local superconducting order parameter is maximal given $\langle {\hat S}_{iz}^2\rangle$\,=\,1. In other words, the superconductivity and magnetism are nonsymbiotic phenomena with competing order parameters giving rise to an inter-twinning, glassiness, and other forms of electronic heterogeneities.
Most likely, superconductivity develops in
the areas of the sample with strongly suppressed spin magnetism down to the lowest temperatures (see, e.g., recent paper\,\cite{Shengelaya}). 
Conventional spin degree of freedom seems to play merely negative effect in high-T$_c$ superconductivity: magnetism is incompatible with optimal high-T$_C$ superconductivity. 
Interestingly, T$^{max}_c$ for doped cuprates anticorrelates with the exchange integral in respective parent compounds\,\cite{Mallett}.

Not long after the discovery of the cuprate superconductors, Johnston\,\cite{Johnston} through an analysis of his measurements of the bulk spin susceptibility found that the temperature dependence of the paramagnetic susceptibility in LSCO and YBCO are consistent with a picture in which both the Pauli contribution $\chi^{Pauli}$ of itinerant hole carriers and the contribution $\chi^{2D}(T)$  of localized spins on the Cu$^{2+}$ ions are present. The very existence of  $\chi^{Pauli}$ which increases with the hole doping implies degeneracy of spin-up and spin-down quasihole states near $E_F$.
The localized spin contribution $\chi^{2D}(T)$ does exhibit dynamic Heisenberg-like intralayer antiferromagnetic fluctuations persisting up to slightly overdoping.  Unconventional temperature and doping behavior of the $\chi^{2D}(T)$ was attributed by Johnston\,\cite{Johnston} to a rapid drop both of the intralayer Cu-Cu exchange integrals and effective Cu spin moments with rising doping. 
However, our spin-pseudospin model Hamiltonian (\ref{Hspin}) with effective exchange integral (\ref{I}) points to a  just another and more justified cause of the puzzle around $\chi^{2D}(T)$, that is a rapid drop of the on-site Cu$^{2+}$ fraction and, accordingly, on-site spin density  $\hat{\rho}^s$.


\section{Going beyond the Zhang-Rice model for the hole C\lowercase{u}$^{3+}$ center and novel order parameters}

The nature of the doped-hole state, or Cu$^{3+}$ center  in the cuprates with nominally Cu$^{2+}$ ions such as La$_2$CuO$_4$ is a matter of great importance in understanding both the mechanism leading to the high-temperature superconductivity and unconventional normal state behavior of the cuprates. 

Well isolated ZR singlet as a ground state of the  Cu$^{3+}$ center in hole doped  cuprates is a leading paradigm in  modern  theories of high-temperature superconductivity.
However, numerous experimental data, in particular, recent magnetic neutron scattering findings (see Refs.\,\onlinecite{Bourges}) and ${}^{6,7}$Li NMR measurements  in La$_2$Li$_{0.5}$Cu$_{0.5}$O$_4$ (Ref.\onlinecite{LaLiCuO}), suggests the involvement of some other physics  which introduces low-lying states into the excitation of the doped-hole state, or competition of conventional ZR state with another electron removal state.
This point was discussed earlier, however, mainly as an interplay between ZR singlet ${}^{1}A_{1g}$ and triplet ${}^{3}B_{1g}$, formed by additional hole going not into $b_{1g}$ state as in ZR singlet, but into $a_{1g}\propto d_{z^2}$ state\,\cite{Eskes}. It is worth noting that ${}^{3}B_{1g}$ state corresponds to a Hund ${}^{3}A_{2g}$ term of two-hole $e_g^2$ configuration of an undistorted CuO$_6$ octahedra. However, later experimental findings for very different insulating cuprates  and theoretical calculations have shown that the energy separation between  the $b_{1g}$($d_{x^2-y^2}$) and $a_{1g}$($d_{z^2}$) orbitals in CuO$_4$ plaquettes is thought to be of the order of 1.5\,eV (see Fig.\,\ref{fig1}), i.e. too large for quasi-degeneracy and effective  vibronic coupling.
More sophisticated version of the non-ZR states was proposed by Varma\,\cite{Varma}, who has proposed that the additional holes doped in the CuO$_2$ planes do not hybridize into ZR singlets, but give rise to  circulating currents on O-Cu-O triangles.

\subsection{The non-ZR A-B-E model and novel IUC order parameters}

\begin{figure}[t]
\includegraphics[width=8.5cm,angle=0]{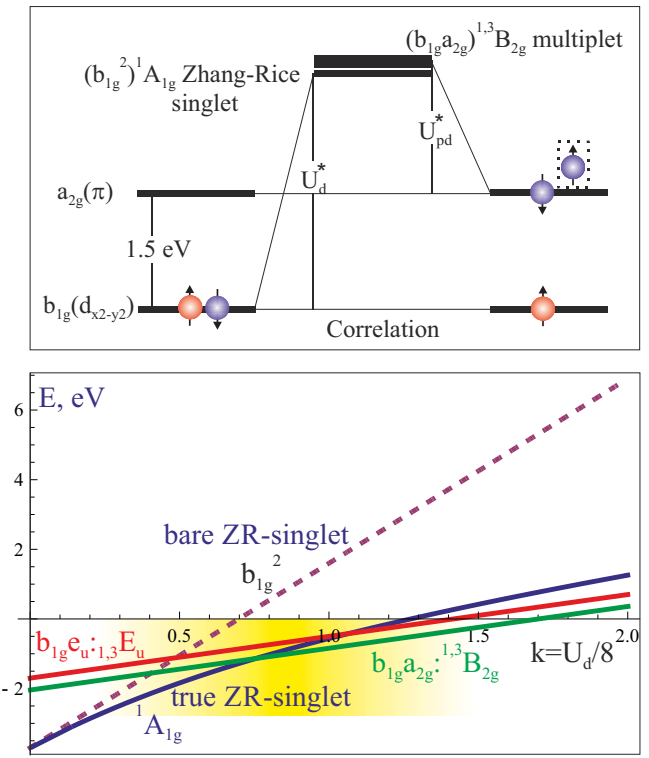}
\caption{(Color online) Top panel: The illustration of the formation of the  valence $^{1}A_{1g}-{}^{1,3}B_{2g}$
multiplet.   Bottom panel: The energies of  bare ZR-singlet, true ZR-singlet, $b_{1g}a_{2g}$, and $b_{1g}e_u$ configurations against the effective "screening" parameter $k=U_d/U_d^0$. Filling points to the quasi-degeneracy region.} \label{fig4}
\end{figure}

Cluster model considerations supported by numerous experimental data point to a competition of conventional hybrid Cu 3d-O 2p $b_{1g}(\sigma)\propto d_{x^2 -y^2}$ state with {\it purely oxygen nonbonding} O 2p$\pi$ states with $a_{2g}(\pi)$ and $e_{ux,y}(\pi) \propto p_{x,y}$ symmetry (see Fig.\,\ref{fig1}, Refs.\onlinecite{NQR-NMR,Moskvin-FNT-11,JETPLett-12} and references therein). 
These orbitals form the energetically  lowest purely oxygen hole states localized on a CuO$_4$ plaquette with the energy $\sim$\,1.5\,-\,2\,eV. At variance with the ZR configuration the "bonding-nonbonding" $b_{1g}^be_u(\pi)$ and $b_{1g}^ba_{2g}(\pi)$ configurations with a full cut off of strong intra-site \emph{d-d} correlations are characterized by a rather small  correlation energy on the order of 1\,ev. Hence the gain in the hole-hole repulsion can compensate the loss in the one-hole energy drawing these configurations to the ZR singlet (see Fig.\,\ref{fig4}).
In other words, these O 2p$_\pi$ states could be as preferable for the localization of additive hole as  the $b_{1g}\propto
d_{x^2-y^2}$ ground state which would result in the instability of the ZR singlet in doped cuprates. Given the same set of different parameters used for the Cu$^{2+}$ cluster calculations (see Fig.\,\ref{fig1}) we calculated\,\cite{Moskvin-FNT-11} the energies of bare $(b_{1g}^b)^{2}$, $b_{1g}^be_u(\pi)$, and $b_{1g}^ba_{2g}(\pi)$ configurations as functions of an effective "screening" parameter $k=U_d/U_d^0$, where $U_d^0=8$\,eV\,\cite{Eskes}, suggesting the same screening for all other Coulomb repulsion parameters (see Fig.\,\ref{fig4}). We see that given $k>$\,0.5 the energy of the  "bonding-nonbonding" $b_{1g}^be_u(\pi)$ and $b_{1g}^ba_{2g}(\pi)$ configurations appears to be lower than the energy of the bare ZR configuration. However, it should be noted that the bare ZR singlet $(b_{1g}^b)^2:{}^1A_{1g}$ can be stabilized due to a sizeable  interaction with $b_{1g}^bb_{1g}^a$ and $(b_{1g}^a)^2$ configurations. 
Despite a large energy separation ($\sim$\,6 and $\sim$\,12\,eV, respectively) all three $(b_{1g})^2$-type configurations interact strongly due to both \emph{d-d} and \emph{p-p} intra-atomic Coulomb coupling. 
The comparison of the energies of the bare ZR-singlet, derived from $(b_{1g}^b)^2$ configuration, and true ZR-singlet (see Fig.\,\ref{fig4}), points to a dramatic role of the configurational interaction (CI)  in the ${}^1A_{1g}$ channel.   It leads to a more $|dp\rangle$ character of the ZR-singlet with a sizeable energy stabilization and strong suppression of the effective correlation energy. Indeed, the step-by-step taking account of the \emph{d-p}$\sigma$ covalency and CI effects results in a strong suppression of the effective correlation energy from bare value $U_{d_{x^2-y^2}^2}$\,=\,8.0 eV to $U_{b_{1g}^2}$\,=\,5.3\,eV and to  $U_{ZR}$\,=\,3.1\,eV, respectively. As a result, we arrive at a quasidegeneracy of the true ZR-singlet ${}^1A_{1g}$ and  $b_{1g}^ba_{2g}$, $b_{1g}^be_u$ configurations taking place in a rather wide range of correlation parameters. 
In other words, instead of a well isolated ground state ZR-singlet for Cu$^{3+}$ centers as in many conventional theoretical approaches, we should consider a complex valent multiplet ${}^1A_{1g}$-$b_{1g}^ba_{2g}(\pi):{}^{1,3}B_{2g}$-$b_{1g}^be_u:{}^{1,3}E_{u}$ (A-B-E -model). The bonding $b_{1g}^b(\sigma)$ -- nonbonding $e_u(\pi)$, $a_{2g}(\pi)$  hole competition reflects a subtle balance between the gain in electron-electron repulsion and  the loss in one-particle energy both affected by a lattice polarization and can be an universal property of a wide group of 2D cuprates. 

Despite a large body of both theoretical and experimental argumentation indirectly supporting the existence of non-ZR multiplets in cuprates their direct experimental probing remains to be highly desirable especially because there are  numerous misleading reports supporting "the stability of simple ZR singlet". For instance,  the authors of the 
photoemission studies on CuO and Bi$_2$Sr$_2$CaCu$_2$O$_{8-\delta}$\,\cite{Tjeng}, have reported that they "are able to unravel the different spin states in the single-particle excitation spectrum of cuprates and show that the top of the valence band is of
pure singlet character, which provides strong support for the existence and
stability of Zhang-Rice singlets in high-$T_c$ cuprates thus
justifying the ansatz of single-band models".  In their opinion "these states are more stable than the triplet states by about 1\,eV".  However, in their photoemission studies they made use of the Cu 2p$_{3/2}(L_{3})$ resonance
condition that allows to detect unambiguously only copper photo-hole states, hence they cannot see the purely oxygen photo-hole $a_{2g}$ and $e_u$ states. 
  
\begin{figure}[t]
\includegraphics[width=6.5cm,angle=0]{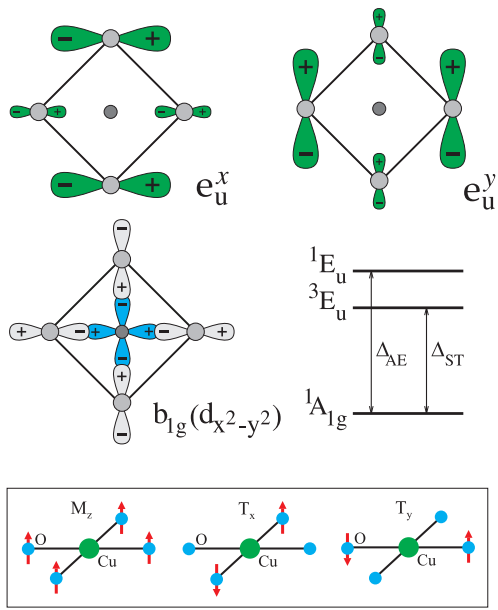}
\caption{(Color online) The term  structure of the actual  valence A-E multiplet for a hole
CuO$_{4}^{5-}$ center together with the single-hole $b_{1g}^{b}$ and
$e_{ux,y}^{b}$  orbitals. Lower panel illustrates the ferromagnetic and toroidal orderings of the oxygen orbital magnetic moments within the CuO$_4$ plaquette.} 
\label{fig5}
\end{figure}

Earlier we have addressed unconventional properties of the non-ZR hole center related to the $^1A_{1g}$-$^{1,3}E_u$ quasi-degeneracy (A-E model)\,\cite{NQR-NMR}. Fig.\,\ref{fig5} shows the term  structure of the actual  valence A-E multiplet together with single-hole basis $b_{1g}^{b}$:
$$
  (|b_{1g}\rangle =c_d |d_{x^2 -y^2}\rangle +c_p|b_{1g}(O2p)\rangle )
$$
and $e_{ux,y}$: 
$$
 (|e_{ux,y}\rangle =c_{\pi} |e_{ux,y}(\pi)\rangle +c_{\sigma}|e_{ux,y}(\sigma)\rangle )
$$
orbitals. 
The $e_{u}$ orbitals could form two  circular current $p_{\pm 1}$-like states, $e_{u\pm 1}$
  with an Ising-like orbital moment 
	$$
	\langle e_{u\pm 1}|l_z|e_{u\pm 1}\rangle = \pm 2c_{\sigma}c_{\pi}\, ,
	$$ 
	which is easily prone to be quenched by a low-symmetry crystal field with formation of two   currentless, e.g., $p_{x,y}$-like $e_{ux,y}$ states. 

Even with neglecting the spin degree of freedom in A-E model we arrive at the eight order parameters for the hole [CuO$_4$]$^{5-}$ center: conventional in-plane electric dipole and quadrupole moments, unconventional Ising-like purely oxygen orbital magnetic moment, and the two-component in-plane purely oxygen orbital IUC toroidal moment (see Fig.\,\ref{fig5}). Microscopically, the effective IUC magnetic toroidal moment can be derived through the local oxygen effective orbital moments as follows:
$$
{\hat {\bf T}}= \beta _{e}\sum _{n=1}^{4} [{\bf R}_{n}\times{\hat {\bf l}}_{n}]\,.
$$
Thus,  the CuO$_4$ plaquette with ($^1A_{1g},^1E_u$) valent multiplet forms an unconventional
magneto-electric center characterized by the eight independent orbital order
parameters. Even this simplified  model predicts  broken  time-reversal ($T$) symmetry, two-dimensional parity ($P$), and basic tetragonal (four-fold Z$_4$) symmetry.
 The situation seems to be more involved, if we take into account spin degree of freedom, in particular, the ${}^1A_{1g}$-${}^{3}E_{u}$ singlet-triplet mixing effects. First of all, such a center is characterized by a true spin S=1 moment being gapped, if the ZR singlet ${}^1A_{1g}$ has the lowest energy. Strictly speaking, for our two hole configuration we should introduce two spin operators: net spin moment ${\hat {\bf S}}={\hat {\bf s}}_1+{\hat {\bf s}}_2$ and  spin operator ${\hat {\bf V}}={\hat {\bf s}}_1-{\hat {\bf s}}_2$ that changes spin multiplicity. It should be noted that the $V$-type order implies an indefinite ground state  spin multiplicity and at variance with $S$-type order is invariably accompanied by an orbital order. The singlet-triplet structure of the A-E multiplet implies two novel types of the spin-orbital order parameters:
spin-dipole parameters $\langle {\hat {\bf V}} {\hat d}_x\rangle$ and $\langle {\hat {\bf V}} {\hat d}_y\rangle$ and spin-toroidal parameters $\langle {\hat {\bf V}} {\hat T}_x\rangle$ and $\langle {\hat {\bf V}} {\hat T}_y\rangle$. Novel ordering does not imply independent V-, d-  or T-type orders. 

Despite a "fragility" \, of the orbital $e_u$-currents with regard to a crystal field quenching  these can produce ferromagnetic-like fluctuations that can explain numerous manifestations of a weak ferromagnetism in different cuprates (see, e.g., Ref.\onlinecite{WM}) and a remarkable observation of a weak magnetic circular dichroism (MCD) in YBa$_2$Cu$_3$O$_{6+x}$\,\cite{Kerr}.  It should be noted that the value of MCD effect does not straightforwardly depends on the value of orbital, or magnetic orbital moment.
  Generally speaking, the hole doped cuprate could be a system with a giant circular
  magnetooptics if we were able to realize the uniform ferromagnetic ordering of the
  orbital $e_u$-currents. It seems likely that  the relative concentration $x_{h}\sim 10^{-4}$ of   circularly polarized $e_u$ holes is enough to provide the same magnitude of MCD as an applied magnetic field of $1$ Tesla\,\cite{MCD}. It is worth noting that the current loop state\,\cite{Varma}, by itself, is incompatible with ferromagnetism and cannot explain the Kerr measurements\,\cite{Kerr}.    




Occurrence of  both orbital toroidal and spin-dipole order parameters point to the hole CuO$_4^{5-}$ centers as polar centers with effective magneto-electric coupling which can provide ferroelectric and magnetoelectric properties for hole-doped cuprates\,\cite{ME}. Interestingly, within the $^1A_{1g},^1E_u$ multiplet the electric dipole moment operator can be coupled with orbital toroidal and magnetic moments by a remarkable magnetoelectric relation\,\cite{NQR-NMR,Moskvin-FNT-11}: 
$$
{\hat d}_x=d_{me}\{{\hat T_y,{\hat M_z}}\}, {\hat d}_y=-d_{me}\{{\hat T_x,{\hat M_z}}\}.
$$

\begin{figure}[t]
\includegraphics[width=6.5cm,angle=0]{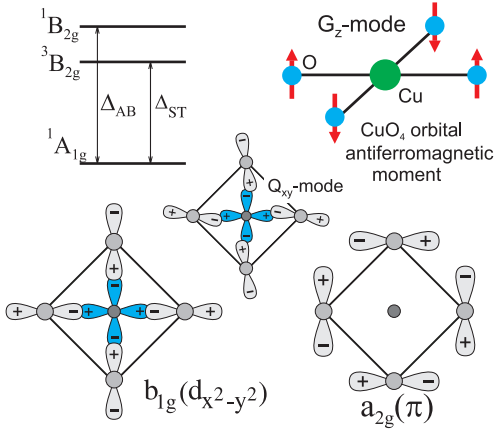}
\caption{(Color online) The term  structure of the actual  valent A-B multiplet for hole
CuO$_{4}^{5-}$ center together with single-hole basis $b_{1g}^{b}$ and
$a_{2g}^{b}$  orbitals. Shown are antiferromagnetic (staggered) ordering of oxygen orbital magnetic moments within CuO$_4$ plaquette ($G_z$-mode) and  quadrupole $Q_{xy}$-mode.} 
\label{fig6}
\end{figure}

Novel effects relate with the $^1A_{1g}$\,-\,$^{1,3}B_{2g}$ quasi-degeneracy (A-B-model)\,\cite{JETPLett-12}.    
Unconventional orbital A-B structure of the hole  CuO$_4$ hole centers 
with the ground state $b_{1g}^2$:${}^1A_{1g}$\,-\,$b_{1g}a_{2g}(\pi)$:${}^{1,3}B_{2g}$ multiplet (see Fig.\,\ref{fig6}) implies several spin, charge, and orbital order parameters missed in the simple ZR model. 
For the orbital quasi-doublet ${}^1A_{1g}$\,-\,${}^{1}B_{2g}$ to be properly described one might make use of a pseudospin formalism with two states ${}^1A_{1g}$ and ${}^{1}B_{2g}$ attributed to $|+\frac{1}{2}\rangle$ and $|-\frac{1}{2}\rangle$ states of a pseudospin $s=\frac{1}{2}$, respectively. Then we introduce three order parameters: $\langle {\hat\sigma}_z\rangle$, $\langle {\hat\sigma}_x\rangle$,  and $\langle {\hat\sigma}_y\rangle$, where ${\hat{\bf \sigma}_i}$ is Pauli matrix. Order parameter $\langle {\hat\sigma}_z\rangle$ defines the symmetry conserving charge density fluctuations within the CuO$_4$ plaquette.
Order parameter $\langle {\hat\sigma}_x\rangle$ defines electric quadrupole moment of $B_{2g}$ symmetry localized on four oxygen sites:
\begin{equation}
Q_{xy}=\sum_{i}{\hat Q}_{xy}(i)=Q_{B_{2g}}\,\langle {\hat\sigma}_x\rangle \,.
\end{equation}
It should be emphasized that the quadrupole moment has an electronic orbital origin (see Fig.\,\ref{fig6}) and has nothing to do with any CuO$_4$ plaquette's distortions or charge imbalance between the density of holes at the oxygen sites. It is worth noting that usually a spontaneous imbalance between
the density of holes at the oxygen sites in the unit cell is related to a $nematic$ order.
Order parameter $\langle {\hat\sigma}_y\rangle$ defines  an antiferromagnetic (staggered) IUC ordering of oxygen orbital  moments localized on four oxygen sites:
\begin{equation}
\langle {\hat G}_z\rangle =\langle {\hat l}_{1z}-{\hat l}_{2z}+{\hat l}_{3z}-{\hat l}_{4z}\rangle =g_L\,\langle {\hat\sigma}_y\rangle \, ,
\end{equation}
where $g_L\approx$\,-1.0, if to make use of estimates of the cluster model\,\cite{Moskvin-FNT-11}.  In other words, maximal value of antiferromagnetic order parameter $G_z$ corresponds to a staggered order of unexpectedly large oxygen orbital magnetic moments $m_z\approx 0.25\,\beta_e$. In contrast with the net orbital moment $M_z$ the $G_z$ order cannot be easily quenched by low-symmetry crystal fields.   
Fig.\,\ref{fig6} shows an illustration of the $G_z$ order in the CuO$_4$ plaquette. 
In fact, both quadrupole moment $Q_{B_{2g}}$ and local antiferromagnetic ordering of oxygen orbital moments $G_z$ do result from the ${}^1A_{1g}$-${}^{1}B_{2g}$ mixing effect, in other words, these are a result of the symmetry breaking. It should be emphasized that the 
$G_z$ order resembles the hotly discussed order of circulating currents,  proposed by Varma\,\cite{Varma}, however, has a more clear physical nature. 

Two unconventional vectorial order parameters are associated with the ${}^1A_{1g}$-${}^{3}B_{2g}$ singlet-triplet mixing effect:  $\langle {\hat{\bf V}} {\hat Q}_{xy}\rangle$ and $\langle {\hat{\bf V}} {\hat G}_{z}\rangle$. It should be noted that corresponding orderings do not imply independent $\langle {\hat{\bf V}} \rangle$,  $\langle {\hat Q}_{xy}\rangle$ or $\langle {\hat G}_{z}\rangle$ orders. Moreover, the $\langle {\hat{\bf V}} {\hat Q}_{xy}\rangle$ and $\langle {\hat{\bf V}} {\hat G}_{z}\rangle$ orders imply all the mean values $\langle {\hat{\bf S}} \rangle$, $\langle {\hat{\bf V}} \rangle$, $\langle {\hat Q}_{xy}\rangle$, $\langle {\hat G}_{z}\rangle$ for CuO$_{4}^{5-}$ center together with their
on-site counterparts such as $\langle {\hat{\bf S}}_i \rangle$, $\langle {\hat Q}_{xy}(i)\rangle$, $\langle {\hat l}_{iz} \rangle$ ($i=Cu,\,O_{1,2,3,4}$) turn into zero, at least in first order on the ${}^1A_{1g}$-${}^{3}B_{2g}$ mixing parameters. 
The most part of novel orbital and spin-orbital order parameters appear to be  strongly  hidden, or the hard-to-detect ones and  can be revealed only by specific experimental technique. 

\subsection{Some experimental manifestation of novel IUC order parameters}
\subsubsection{A-B model and the  G$_z$-type IUC order in La$_2$Li$_{0.5}$Cu$_{0.5}$O$_4$}

A unique opportunity to study  isolated Cu$^{3+}$ centers and non-ZR IUC order parameters without the confounding contributions of the nearest neighbor  antiferromagnetically correlated CuO$_4$ clusters is provided in La$_2$Li$_x$Cu$_{1-x}$O$_4$ at $x=0.5$\,\cite{Demazeau}. At this composition the Li and Cu ions form an ideally ordered superlattice\,\cite{Attfield,Moshopoulou}  in which all  Cu$^{3+}$ ions are surrounded by four in-plane Li ions (1s$^2$, closed shell electronic configuration) and thus create weakly coupled, almost isolated CuO$_4$  clusters. 

First indication of a breaking of isolation for the ZR ground state in this compound was obtained by Yoshinari {\it et al}.\,\cite{Yoshinari}: 
their analysis of the temperature dependence of the ${}^{63,65}$Cu  NQR (nuclear quadrupole resonance) relaxation rates unambiguously evidenced that the singlet state has a 130\,meV gap to magnetic excitations. In other words, it appeared the energy of the excited spin-triplet state relative to the ground state is radically smaller than predicted by Zhang and Rice\,\cite{ZR} and many other authors.

Both $^{63,65}$Cu NQR\,\cite{Yoshinari},  $\mu$SR studies\,\cite{mu-LaLiCuO}, and especially recent low-temperature ${}^{6,7}$Li NMR (nuclear magnetic resonance)\,\cite{LaLiCuO} measurements  in La$_2$Li$_{0.5}$Cu$_{0.5}$O$_4$
revealed  significant magnetic fluctuations and did  suggest  some kind of  quasi-degeneracy in the valence state of the hole centers with 
a competition of the conventional ZR state with another low-lying state(s).
Direct information about these states can be retrieved from the low-temperature neutron structural data\,\cite{Attfield} on La$_2$Li$_{0.5}$Cu$_{0.5}$O$_4$ system which revealed a $B_{2g}$-type (rectangular) distortion of both CuO$_4$ and LiO$_4$ plaquettes with acute in-plane O-M-O bond angles 86$^{\circ}$ and 87$^{\circ}$, respectively.
This finding which was supported later by the electron diffraction studies\,\cite{Moshopoulou} cannot be reconciled with the concept of the well isolated ZR singlet but agrees with a static pseudo-Jahn-Teller(PJT)-effect induced by a vibronic coupling of the $^{1}A_{1g}$ ZR singlet with a nearby ${}^{1}B_{2g}$ singlet:
$$
V_{PJT}=v\,{\hat\sigma}_x\,u_{B_{2g}} \, ,
$$ 
where $u_{B_{2g}}$ is a $B_{2g}$-type symmetry combination of the oxygen displacements in CuO$_4$ plaquette, ${\hat\sigma}_x$ is Pauli matrix on the $^{1}A_{1g}$\,-\,${}^{1}B_{2g}$ doublet (see above), $v$ is a vibronic coupling constant.  
The vibronic mixing of the ${}^{1}A_{1g}$ and ${}^{1}B_{2g}$ terms gives rise to a two-well adiabatic potential with two types of the bond-bending $B_{2g}$ distortion of the CuO$_4$ plaquette: $B_{2g}^x$ and $B_{2g}^y$, respectively. Their in-plane long-range ferrodistortive ordering with a rectangular distortion both of CuO$_4$ and LiO$_4$ plaquettes gives rise to the orthorhombic $Ammm$ structure\,\cite{Attfield,remark}.

However, as we see above, the $^{1}A_{1g}$\,-\,${}^{1}B_{2g}$ doublet  would inevitably  be characterized by one more IUC order parameter, that is unconventional antiferromagnetic  $G_z$ order of the oxygen orbital/magnetic moments localized on four oxygen sites. In other words,  
 a quantum disordered ferrodistortive fluctuating $Ammm$ order in 2D structure of the (CuLi)O$_2$ planes\,\cite{Attfield} would be accompanied by unconventional oxygen orbital antiferromagnetic fluctuations ($Ammm-G_z$ mode).
The staggered $G_z$-type IUC orbital magnetic fluctuations do not produce local fields on central $^{63,65}$Cu nuclei and can be seen only by $^{17}$O  or $^{6,7}$Li nuclei  rather than by $^{63,65}$Cu nuclei.
This findings did stimulate the ${}^{6,7}$Li NMR measurements in La$_2$Li$_{0.5}$Cu$_{0.5}$O$_4$ in a wide temperature range  down to T\,=\,2\,K\,\cite{LaLiCuO}. We have observed a dramatic low-temperature evolution of the ${}^{6,7}$Li NMR lineshape and the spin-lattice relaxation  (SLR) rate  indeed pointing to a gradual slowing down of some specific $magnetic$ order parameter's fluctuations without distinct signatures of a phase transition down to T\,=\,2\,K. An unique opportunity to distinguish between the magnetic and charge distribution caused phenomena are provided by a comparison of NMR lineshapes for $^6$Li and $^7$Li nuclei: due to the very low quadrupole moment of $^6$Li nuclei (${}^6Q$\,=\,-0.0008\,barn) as compared to $^7$Li  (${}^7Q$\,=\,-0.045\,barn) the $^6$Li NMR spectra are almost free from any quadrupole effects.

The ${}^{6,7}$Li NMR data completely agrees with predictions of the A-B model with regard the $G_z$-type IUC orbital magnetic fluctuations and evidences a step-by-step condensation of the ferrodistortive lattice-orbital $Ammm-G_z$ mode.
Furthermore, the $Ammm-G_z$ mode fluctuations also allows for explaining the low-temperature $^{63,65}$Cu NQR data\,\cite{Yoshinari}. Indeed,  
well developed  $B_{2g}$-type symmetry distortions of the CuO$_4$ plaquettes do explain the domination of quadrupolar mechanism for the $^{63,65}$Cu nuclear relaxation below 170\,K with unusual temperature dependence of $T_1^{-1}$ strongly deviating from $T_1^{-1}\propto T^2$  characteristic of a phononic nuclear spin relaxation mechanism. A weak yet distinct intrinsic {\it magnetic orbital} contribution to  $T_1^{-1}$ having the same T dependence\,\cite{Yoshinari} can be naturally explained, if we take into account weak supertransferred magnetic hyperfine interactions between neighboring CuO$_4$ plaquettes. It is worth noting that the distribution of $T_1^{-1}$'s\,\cite{Yoshinari} demonstrates that not all Cu sites are equivalent and the local crystal structure seems to vary on a nanoscopic length scale thus also supporting the picture of the $Ammm-G_z$ mode fluctuations. 
It should be noted that the A-B-E quasi-degeneracy of the valence multiplet does not necessarily produce large positive magnetic susceptibility for the hole CuO$_4^{5-}$ centers. Indeed, given the ZR singlet as a ground state we arrive at a dominant contribution of the fluctuations induced by the A-B or A-E "linear" \,mixing which does not produce the net spin or orbital magnetic moment. This agrees with the negative susceptibility observed for La$_2$Li$_{0.5}$Cu$_{0.5}$O$_4$ down to very low temperatures T\,$\sim$\,10\,K\,\cite{Rykov} that was considered earlier to be a strong argument in favor of the well isolated spin singlet ZR ground state.
It should be added that the A-B-E model suggests even two candidate spin triplets, ${}^{3}E_{u}$ or ${}^{3}B_{2g}$, for  the magnetic excitation at 0.13\,eV observed in the ${}^{63,65}$Cu  NQR studies of La$_2$Li$_{0.5}$Cu$_{0.5}$O$_4$\cite{Yoshinari}.

At present, there are no published NMR or ZF-$\mu$SR studies which revealed signatures of static G$_z$ type mode in cuprates (see Ref.\,\onlinecite{Bourges} and references therein). The failure to detect orbital-like magnetic order of the kind observed by spin-polarized neutron diffraction\,\cite{Bourges} surely indicates that the local fields are rapidly fluctuating outside the $\mu$SR or NMR time window. In this regard the $^{6,7}$Li NMR measurements in La$_2$Li$_{0.5}$Cu$_{0.5}$O$_4$ can be addressed as a first indication of a quasi-static G$_z$-type mode realized in cuprates.

\subsubsection{Manifestation of the  G$_z$-type IUC order in the spin-polarized neutron diffraction measurements}


Theoretical cross section formulae for the scattering of neutrons by magnetic ions in crystals were first derived in 1953 by Trammell\,\cite{Trammell}, using the traditional Condon and Shortley formalism\,\cite{Condon}. However, in order to use the Trammell formulae it is necessary to evaluate two different sets of radial integrals, one set for the orbital and one set for the spin contribution. Later on Johnston\,\cite{Johnston-69} based on the formalism of Racah algebra\,\cite{Varshalovich,Racah} has given an elegant calculation of the cross section  by directly calculating the matrix elements. In contrast to the analysis of Trammell, Johnston's expression for the cross section contains only the one set of radial integrals $j_k$ for the orbital and  spin contribution.
The apparent difference between the results of the two calculations arises because Trammell chose to calculate not the actual
matrix elements appearing in the cross section but an intermediate function so constructed as to retain the same simple structure for the cross section as in the spin-only case. However, this intermediate function is not uniquely defined and any function proportional to the scattering vector  may be added to it without altering the expression for the cross section. 

Earlier in Ref.\,\onlinecite{JETPLett-12} we have made an attempt to calculate the magnetic neutron diffraction amplitude for the neutron coupling with the spin-orbital A-B-E multiplet making use of the Trammel technique. However, the technique  has brought us to some inconsistencies, especially for main contribution that is of the $G_z$-type IUC order.  Hereafter, we revisit the calculation in frames of the Johnston technique following the review article by Lovesey and Rimmer\,\cite{Lovesey}.

The magnetic neutron diffraction amplitude is determined by matrix elements of a vector operator\,\cite{Lovesey}
\begin{equation}
	\hat {\bf D}=\sum_{}e^{i\left( {\bf k}\cdot{\bf r}\right)}\left({\bf s}-\frac{1}{k^2}\left[{\bf k}\times{\bf \nabla}\right]\right) \, ,
	\label{D}
\end{equation}
 to be a sum of spin and orbital contributions. 
 In (\ref{D}), the sum runs over all unpaired electrons in the material of interest, $\bf k$ is the scattering
wavevector, and ${\bf r}$ and ${\bf s}$, respectively, are position and spin operators.

With respect to the basis of the $2l+1$ atomic $nlm$-functions the orbital operator in (\ref{D}) can be replaced by equivalent operator as follows\,\cite{Lovesey}: 
$$
	\hat {\bf D}^L=-\frac{1}{k^2}e^{i\left( {\bf k}\cdot{\bf r}\right)}\left[{\bf k}\times{\bf \nabla}\right]_q \equiv  
	\frac{1}{k}\sum_{K^{\prime}K^{\prime\prime}}B_{K^{\prime}K^{\prime\prime}}
	$$
	\begin{equation}
	\left(\left\langle j_{K^{\prime}-1}\right\rangle_{nl} + \left\langle j_{K^{\prime}+1}\right\rangle_{nl}\right) 
	\left[V^{K^{\prime}}(l)\times C^{K^{\prime\prime}}({\bf k})\right]^1_q \, ,
	\label{DL}
\end{equation} 
where $V^{K^{\prime}}(l)$ is an irreducible tensor operator with matrix
$$
\left\langle lm|V^{K^{\prime}}_{Q^{\prime}}(l)|lm^{\prime}\right\rangle = (-1)^{l-m}\left(
\begin{array}{ccc}l&K^{\prime}&l
\\
-m&Q^{\prime}&m^{'}\end{array}\right) \, ,
$$

$\left[V^{K^{\prime}}(l)\times C^{K^{\prime\prime}}({\bf k})\right]^1_q$ is the rank 1 irreducible tensorial product, $C^{K^{\prime\prime}}({\bf k})$ is tensorial spherical harmonics, $\left\langle j_{K^{\prime}\pm 1}\right\rangle_{nl}$ is a radial mean value of the spherical Bessel function;
$$
B_{K^{\prime}K^{\prime\prime}}=i^{K^{\prime}+1}[l,K^{\prime},K^{\prime\prime}]\sqrt{(2l+2)(2l+3)}\left(\begin{array}{ccc}1&K^{\prime}&K^{\prime\prime}
\\
0&0&0\end{array}\right)
$$
\begin{equation}
 \left(\begin{array}{ccc}l&K^{\prime}&l+1
\\
0&0&0\end{array}\right) \left\{\begin{array}{ccc}1&1&1
\\
K^{\prime\prime}&K^{\prime}&K^{\prime}\end{array}\right\}\left\{\begin{array}{ccc}l&1&l+1
\\
K^{\prime}&l&K^{\prime}\end{array}\right\}\, ;
\end{equation}
$[l,...]$\,=\,$(2l+1)\cdot ...$, $K^{\prime}$ is an odd number, 1\,$\leq$\,$K^{\prime}$\,$\leq$\,$2l-1$;  $K^{\prime\prime}$ is an even number, $K^{\prime\prime}$=$K^{\prime}\pm 1$. Everywhere we make use of a standard notation for Wigner coefficients and 6$j$-symbols.

Tensor operator $V^{1}(l)$ can be expressed through orbital moment operator as follows: 
\begin{equation}
	V^{1}_q(l)=[l(l+1)(2l+1)]^{-\frac{1}{2}}\,{\hat l}_q \, .
\end{equation}
 For $np$-electrons $K^{\prime}$\,=\,1, $K^{\prime\prime}$\,=\,0 or 2 and Exp.(\ref{DL}) yields an anisotropic link of vector operator $\hat {\bf D}^L$ with orbital moment operator, e.g., for ${\bf l}\parallel O_z$
\begin{eqnarray}
	{\hat D}^L_i\equiv -\frac{\left\langle j_{0}\right\rangle_{np} + \left\langle j_{2}\right\rangle_{np}}{k\sqrt{2}}
\left\{\begin{array}{c}e_xe_z, i=x 
\\
e_ye_z, i=y
\\
(e_z^2-1), i=z 
\end{array}\right\} {\hat l}_z,
\end{eqnarray}
where ${\bf e}$\,=\,${\bf k}/k$ is a unit scattering vector.

Now we can apply the theory to the coupling with the A-B doublet of the Cu$^{3+}$ center. First of all we should write out the expression for 
the $\hat {\bf D}^L$ operator relevant for the two-hole [CuO$_4$]$^{5-}$ cluster with the $G_z$-type order of oxygen orbital moments:   
$$
\hat {\bf D}^L=-\frac{1}{k^2}\sum_{n,\nu}e^{i\left( {\bf k}\cdot({\bf R}_n+{\bf r}_{n\nu})\right)}\left[{\bf k}\times{\bf \nabla_{n\nu}}\right]
$$
where sum runs on the four oxygen ions ($n$\,=\,1-4) and two holes ($\nu$\,=\,1,2).


An easy algebra allows us to show that with respect to the A-B basis the orbital operator $\hat {\bf D}^L$ can be replaced by equivalent operator as follows:
\begin{equation}
\hat {\bf D}^L\equiv {\bf d}{\hat G}_z=g_L{\bf d}{\hat \sigma}_y \, ,	
\end{equation}
where  
\begin{eqnarray}
	d_i =-\frac{1}{2}(\cos k_xl-\cos k_yl) \\ \nonumber
	\frac{\left\langle j_{0}\right\rangle_{np} + \left\langle j_{2}\right\rangle_{np}}{k\sqrt{2}}
\left\{\begin{array}{c}e_xe_z, i=x 
\\
e_ye_z, i=y
\\
(e_z^2-1), i=z 
\end{array}\right\} \, ,
\label{d}
\end{eqnarray}
where $l=R_{CuO}\approx a/2$. Note that all the components $d_i$ are expressed in terms of the same radial integrals.

The G$_z$-type IUC order in the cuprate CuO$_2$ planes is believed\,\cite{JETPLett-12} to be responsible for an unusual translational-symmetry preserving antiferromagnetic order which was recently revealed by the spin-polarized neutron diffraction in the pseudogap phase of several hole-doped high-T$_c$ cuprates, YBCO, LSCO, Bi2212, and Hg1201\,\cite{Bourges}.
The magnetic scattering appears to be the largest at  Bragg indices (H,K,L)=(1,0,L),  (0,1,L) for any integer L value along c* and the largest magnetic intensity is observed for L=0. At large $|{\bf Q}|$ the magnetic form factor is expected to considerably reduce the signal. Accordingly, measurements at the Q=(2,0,1) reflection show no magnetic scattering. Effective magnetic moment $\sim$\,0.1\,$\mu_B$ seen by neutrons appears to be tilted with respect to the c-axis with the tilting angle  $\theta$\,$\approx$\,45\,$\pm$\,20$^{\circ}$.

All these puzzleties can be naturally explained to be a result of the neutron coupling with the hidden G$_z$-type IUC order in the cuprate CuO$_2$ planes. First, the ${\bf Q}$-dependent factor $(\cos k_xl-\cos k_yl)$ in (\ref{d}) points to the Bragg indices (1,0,L) and  (0,1,L) as the most preferable for the magnetic scattering. Second, as we see, in contrast with the spin moment the oxygen orbital moments directed perpendicular to the  CuO$_4$ plaquette in the $G_z$  mode   induce an effective magnetic coupling both with $z$- and $x$-, and/or $y$-components of the neutron spin: neutrons see the effective orbital magnetic moments to be tilted in ${\bf c}^*$-${\bf q}$ plane. Furthermore, any appeal to "an effective magnetic moment seen by neutrons"\, seems to be incorrect in this case  if only because of the length of the "moment"\, would not be invariant being dependent on the direction of the scattering vector ${\bf k}$: indeed, $|{\bf d}|^2\propto (1-e_z^2)$. Nevertheless, we can introduce "an effective tilt angle"\, $\theta$:
\begin{equation}
	cos^2\theta = \frac{d_z^2}{|{\bf d}|^2}=(1-e_z^2) \,.
\end{equation}
  The effective tilt angle turns into zero for Bragg vectors such as ${\bf q}=(010)$ or $(100)$, while e.g. for Bragg vector  $(101)$ in the bilayer system Y123\,\cite{Bourges}($|{\bf Q}|$\,=\,1.71\,$\AA^{-1}$, $Q_L$\,=\,0.54\,$\AA^{-1}$) $\theta$\,$\approx$\,18$^{\circ}$,
for Bragg vector  $(103)$ in the	bilayer system Bi2212 ($|{\bf Q}|$\,=\,1.75\,$\AA^{-1}$, $Q_L$\,=\,0.61\,$\AA^{-1}$) $\theta$\,$\approx$\,20$^{\circ}$. 
Experimental data available for several cuprates\,\cite{Bourges} agrees with the predicted $L$-dependence, however,  usually yields larger values for $\theta$: e.g. the most accurate experiment on YBa$_2$Cu$_3$O$_{6.6}$\,\cite{Bourges} yields $\theta\approx$\,35$^{\circ}$ at (1,0,0) and $\theta\approx$\,55$^{\circ}$ at (1,0,1). At the same time, the most recent data for Bragg vector  $(103)$ in Bi2212\,\cite{Bourges} yields $\theta$\,=\,20\,$\pm$\,20$^{\circ}$ in a full accordance with our estimates. 
Some disagreement between model predictions and experiment for YBCO can be seemingly explained as a result of a buckling effect that is of a nonparallelism of CuO$_4$ plaquettes and CuO$_2$ planes, yielding a tilt of about $14^{\circ}$.

 

To explain experimental data\,\cite{Bourges} we do not need
to engage the spin-orbital coupling,  quantum corrections\,\cite{Varma} or
orbital currents involving the apical oxygens\,\cite{Weber} as the measured polarization effects can be explained with the  locally staggered oxygen orbital moments orthogonal to the CuO$_2$ planes. It is worth noting that the $G_z$-type ordering preserving the translational symmetry cannot be detected in the polarized elastic neutron scattering measurements performed  at the Bragg scattering wave vector such as ${\bf q}=(11L)$ that does explain earlier unsuccessful polarized neutron reports\,\cite{Lee}.

Oxygen orbital moments must inevitably generate local magnetic fields, first of all it concerns a giant $\sim$ 1\,T field (given oxygen magnetic moment of $\sim$\,0.1\,$\mu_B$), directed perpendicular to the CuO$_2$ plane,  at the oxygen nuclei. 
However, the $^{17}$O NMR data on very different cuprates \,\cite{17O-NMR} do not reveal signatures of static G$_z$ type
mode. At present, there are no published $^{63,65}$Cu or $^{17}$O NMR studies which give clear results concerning the existence or absence of fields of the predicted magnitude in YBCO, La-214, Hg1201 or Bi2212. 
The $G_z$-type orbital magnetic order, as any other moment patterns which have reflection symmetry across the Cu-O-Cu bonds would generate a zero magnetic field on yttrium and barium sites in YBa$_2$Cu$_3$O$_{6+\delta}$, YBa$_2$Cu$_4$O$_{8}$, Y$_2$Ba$_4$Cu$_7$O$_{15-\delta}$,  thus making direct $^{89}$Y and $^{135,137}$Ba NQR/NMR  methods as "silent local probes"\, despite their pronounced sensitivity for weak local magnetic fields. This reconciles  the "non-observance"\, results obtained by ${}^{89}$Y NMR in superconducting Y$_2$Ba$_4$Cu$_7$O$_{15-\delta}$ and   
  ${}^{135,137}$Ba NQR in superconducting YBa$_2$Cu$_4$O$_{8}$\,\cite{Strassle} with neutron scattering results\,\cite{Bourges}.
The ZF-$\mu$SR measurements in  YBa$_2$Cu$_3$O$_{6+\delta}$ and La$_{2-x}$Sr$_x$CuO$_4$\,\cite{Sonier} have also found no evidence for the onset of magnetic order at
the pseudogap temperature T$^*$. The NMR and $\mu$SR experiments clearly rule  static G$_z$ type order out. 
The failure to detect orbital-like magnetic order of the kind observed by spin-polarized neutron diffraction surely indicates that the local fields are rapidly fluctuating outside the $\mu$SR or NMR time window or the order is associated with a small minority phase that evolves with hole doping\,\cite{Sonier}.

Above we have addressed the G$_z$-type IUC order for the Cu$^{3+}$ centers, however, as we argued in Sec.II the CuO$_4$ centers in doped cuprates should be described by mixed valence superpositions (\ref{function}).  These are described both by novel orbital order parameters such as the G$_z$-type IUC order through the Cu$^{3+}$ term and by conventional spin order parameter through the Cu$^{2+}$ term. 
In other words, both these competing orders should be seen by neutrons. Obviously, conventional spins in doped cuprates, as in parent compounds, do prefer the in-plane orientation thus making an additional if not leading contribution to the mean tilting angle $\theta$.

We should mention once again that the $G_z$-type IUC order is related with the non-ZR hole Cu$^{3+}$ centers, hence its manifestation would strongly differ in hole- and electron-doped cuprates. We hope that this point will stimulate further experimental studies on the both types of cuprates.    

\section{Conclusions}

 We have presented an unified approach to the description of the variety of the local intra-unit-cell (IUC) order parameters determining a low-energy physics in cuprates. Central point of the model implies the on-site Hilbert space  reduced to only three effective valence centers CuO$_4^{7-,6-,5-}$ (nominally Cu$^{1+,2+,3+}$) and the occurrence of unconventional on-site quantum superpositions of the three valent states characterized by different hole occupation: $n_h$=0,1,2 for Cu$^{1+,2+,3+}$ centers, respectively, different conventional spin: s=1/2 for Cu$^{2+}$ center and s=0 for Cu$^{1+,3+}$ centers, and different orbital symmetry:$B_{1g}$ for the ground states  of the  Cu$^{2+}$ center and  $A_{1g}$   for the Cu$^{1+,3+}$ centers, respectively.  
To describe the diagonal and off-diagonal, or quantum local charge order we develop an S=1 pseudospin model with a non-Heisenberg effective Hamiltonian that provides a physically clear description of "the myriad of phases"\,  from a bare parent  antiferromagnetic insulating phase to a Fermi liquid  in  overdoped  cuprates.  Different orbital symmetry of the ground states  for  Cu$^{2+}$ and  Cu$^{1+,3+}$ does result in a spontaneous orbital symmetry breaking accompanying the formation of the on-site mixed valence superpositions with emergence of the IUC orbital nematic order parameter of the $B_{1g}=B_{1g}\times A_{1g}$ ($\propto d_{x2-y2}$) symmetry.
Conventional spin density $\rho_s$ for mixed valence superpositions can vary inbetween 0 and 1 in accordance with the weight of the Cu$^{2+}$ center in the superposition. We show that the superconductivity and spin magnetism are nonsymbiotic phenomena with competing order parameters. 

Furthermore we argue that instead of a well-isolated Zhang-Rice (ZR) singlet $^1A_{1g}$ the ground state of the hole  Cu$^{3+}$ center in cuprates should be described by a complex $^1A_{1g}$-$^{1,3}B_{2g}$-$^{1,3}E_u$ multiplet, formed by a competition of conventional hybrid Cu 3d-O 2p $b_{1g}(\sigma)\propto d_{x^2 -y^2}$ state and {\it purely oxygen nonbonding} O 2p$\pi$ states with $a_{2g}(\pi)$ and $e_{ux,y}(\pi)$ symmetry. In contrast with inactive ZR singlet we arrive at several novel competing IUC orbital and spin-orbital order parameters, e.g., electric dipole and quadrupole moments, Ising-like net orbital magnetic moment,  orbital toroidal moment, intra-plaquette's staggered order of Ising-like oxygen orbital magnetic moments. As a most impressive validation of the non-ZR model we explain fascinating results of recent neutron scattering measurements that revealed novel type of the IUC magnetic ordering in pseudogap phase of several hole-doped cuprates.

It is worth noting that the spin-pseudospin system in 2D cuprates is prone to a topological phase separation with formation of novel types of a spin-charge order, including (multi)skyrmionic structures\,\cite{EBHM}.

As a whole, our approach allows to shed light on the interplay between d-wave superconductivity, IUC electronic nematicity, IUC magnetic order and other electronic states such as an incipient CDW/SDW order.

I wish to thank Philippe Bourges for stimulating discussions.
The work was supported by the Government of the Russian Federation  Program  02.A03.21.0006 and the  RFBR grant No. 12-02-01039.


\end{document}